\documentstyle[emulateapj,onecolfloat]{article}

\newcommand{\beq}{\begin{equation}}
\newcommand{\eeq}{\end{equation}}
\def\beqa{\begin{eqnarray}}
\def\eeqa{\end{eqnarray}}
\newcommand{\lsim}{\lesssim}
\newcommand{\gsim}{\gtrsim}
\newcommand{\lya}{Ly~$\alpha$\ }
\newcommand{\ba}{{\rm b}}
\newcommand{\mat}{{\rm m}}
\begin{document}
\twocolumn[
\title{Delayed Recombination}

\author{
P. J. E. Peebles$^{1}$, 
S. Seager$^{2}$, and Wayne Hu$^{2}$}
\affil{
{}$^1$Joseph Henry Laboratories, Princeton University,
Princeton, NJ 08544\\
{}$^2$Institute for Advanced Study, Olden Lane, Princeton, NJ
08540\\ 
}
 
\begin{abstract}
Under the standard model for recombination of the primeval
plasma, and the cold dark matter model for structure formation, 
recent measurements of the first peak in the angular power
spectrum of the cosmic microwave background temperature indicate
the spatial geometry of the universe is nearly flat. If sources
of \lya resonance radiation, such as stars or active galactic
nuclei, were present at $z\sim 1000$ they would delay recombination, 
shifting the first peak to larger angular scales, and producing 
a positive bias in this measure of space curvature. It can be 
distinguished from space curvature by its suppression of the
secondary peaks in the spectrum. 
\end{abstract}

]

\section{Introduction}
The measurements of the anisotropies of the cosmic
microwave background (the CMB) offer extraordinarily powerful
tests of the relativistic Friedmann-Lema\^\i tre cosmological
model and the nature of the early stages of cosmic structure
formation (e.g. \cite{Junetal96} 1996). 
Indeed, the recent detection of the first peak in the angular 
power spectrum of the CMB temperature indicates space
is close to flat (\cite{Miletal99} 1999; 
\cite{Meletal99} 1999; \cite{deBer00} 2000). The interpretation
is quite indirect, however so we must seek
diagnostics for possible complications. 

There are relatively few physical effects that can {\it increase} the
angular scale of the first peak.  Because of this fact, the observed 
large angular scale of the peak is believed to strongly disfavour open universes.  
Of the fundamental cosmological parameters, only a Hubble constant
well in excess of observations can substantially increase the scale
of the peak in an open or flat universe (\cite{HuSug95} 1995).   

One possibility is that
some process at redshift $z\sim 1000$ delayed recombination of 
the primeval plasma. This would increase the sound horizon at last
scattering and decrease the angular size distance to last
scattering, moving the first peak of the CMB temperature
fluctuation spectrum to smaller angular wavenumber
(\cite{HuWhi96} 1996; \cite{WelBatAlb99} 1999), 
and biasing the measure of space
curvature to a too large (more positive) apparent value. 
Another consequence of delayed recombination is that it would
suppress the secondary peaks due to an increase in the time available
for acoustic oscillations to dissipate  (\cite{Sil68} 1968; \cite{HuWhi96} 1996).

Because the first peak is not observed to have suffered substantial
dissipation, recombination in the delayed model must be rapid
compared with the cosmological expansion rate. 
Thus if recombination were delayed by ionizing radiation from decaying
dark matter (e.g. \cite{SarCoo83} 1983; 
\cite{Sciama} 1991; \cite{Elletal92} 1992) or evaporating
primeval black holes (\cite{pbh} 1987), or by thermal energy 
input from cosmic string wakes (\cite{WelBatAlb99} 1999), the 
source would have to terminate quite abruptly and well before 
$z\sim 100$ when the universe starts to become optically thin even if
the baryons are fully ionized. 

Here we consider a picture that more naturally allows rapid
recombination: sources of radiation at $z\sim 1000$ that produce
many more photons in the resonance \lya line than ionizing
photons, in the manner of a quasar. The \lya photons would
increase the population in the principal quantum number $n=2$
levels of atomic hydrogen, increasing the rate of photoionization 
from $n=2$ by the CMB. Since the rate of thermal photoionization 
from $n=2$ varies rapidly with redshift at $z \lsim 1000$, the
delayed recombination is rapid, and the residual ionization can
be small enough that the optical depth for Thomson scattering
after recombination is well below unity. Thus the height of the
first peak in the spectrum of CMB temperature fluctuations is
little affected by the delayed recombination. The shift in the
angular wavenumber at the first peak is modest also, even
if early sources produce many \lya photons per baryon, but the shift can
be considerably larger than the projected precision of the
measurements. Thus it is fortunate that we seem to have an unambiguous 
diagnostic for delayed recombination in the suppression
of the secondary peaks. 

After this work was substantially complete we learned that 
the recent measurements by the BOOMERanG experiment in fact favor
a substantial suppression of the second peak (\cite{deBer00} 2000).
We emphasize that there are many other ways to account for this
effect, and that there is no evidence for the early source 
of \lya radiation assumed in our picture. Within the conventional
adiabatic CDM model the recombination history is well understood
(\cite{SeaSasSco00} 2000 and earlier references therein), so
there is good reason to expect the residual fluctuations in the
CMB may be related to the cosmology in a simple and computable
way. But it is good science to bear in mind the possibility that
Nature is more complicated than our ideas, a rule that has
particular force in cosmology because of the limited empirical
basis. 

\begin{figure}
\plotone{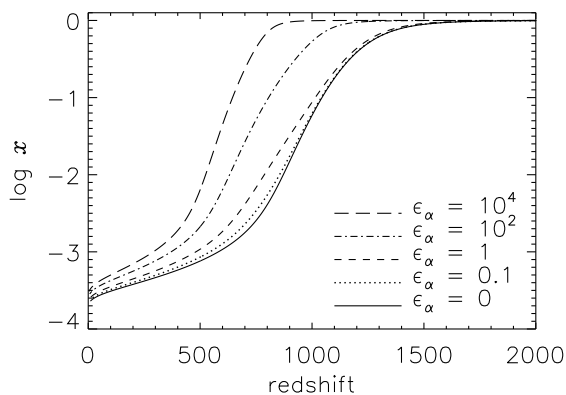}
\caption{The ionization fraction, $x$, as a function of redshift
for $\epsilon_{i}=0$ and various values of $\epsilon_{\alpha}$.}
\end{figure}

\section{The Model}

Since the early source of radiation is purely conjectural, a
simple model is appropriate. We assume the rate of production of
\lya resonance photons per unit volume (in excess of those
produced by the primeval plasma) is
\beq
dn_\alpha /dt = \epsilon _{\alpha} n_{\rm H}H(t),
\label{eq:source}
\eeq
where $n_{\rm H}$ is the number density of hydrogen nuclei, 
$H(t)=\dot a/a$ is the expansion rate, and $\epsilon _{\alpha}$ is a free parameter. 
The computation of the ionization history by \cite{SeaSasSco00} (2000) 
is readily adjusted to take account of this new
source term if it is approximately homogeneous. 
The results in Figure~1 assume the Hubble parameter
is $h=0.7$ in units of $100$ km~s$^{-1}$~Mpc$^{-1}$, the baryon
density parameter is $\Omega _\ba h^2=0.02$ 
(\cite{Tytler} 2000) and the
total matter density parameter is $\Omega _\mat= 0.25$. The
ionization history is quite similar at $\Omega _\mat=0.4$. 

The optical depth for Thomson scattering subsequent to
redshift $z$ is
\beq
\tau = 63 x \left( {1-Y_{p} \over 0.76} \right)
          \left( { \Omega_{\ba} h^{2} \over 0.02 } \right)
	  \left( { \Omega_{\mat}h^{2} \over 0.122} \right)^{-1/2}
	  \left( z \over 10^{3} \right)^{3/2},
\label{eq:tau}
\eeq
if the fractional ionization is constant at $x$ at redshift
less than $z$.  
Figure~1 shows that the delayed recombination
through the redshift $z_*$ where $\tau$ passes through unity is about as
fast relative to the expansion rate as in the standard
recombination model.  Furthermore the decrease in $z_{*}$ itself
is approximately 
\begin{equation}
{z_*(\epsilon_{\alpha}) \over z_*(0)} = (1+3\epsilon_{\alpha})^{-0.042} \,.
\label{eqn:shift}
\end{equation}

\begin{figure}
\plotone{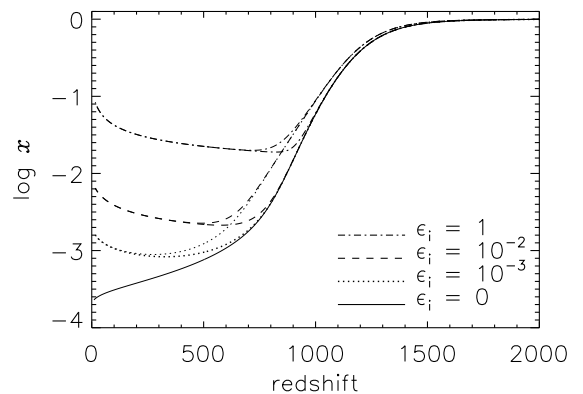}
\caption{The effect of new sources of ionizing photons on the
recombination history. For each value of $\epsilon_{\rm i}$ the
upper curve is for $\epsilon_{\alpha}=1$ and the lower curve
for $\epsilon_{\alpha}=0$.}
\end{figure}

Sources of \lya photons may also produce ionizing radiation.
Since the rate of recombination of 
fully ionized baryons at the cosmic mean density is faster than
the rate of expansion the ionization, $x$ may be approximated by
the equilibrium equation   
\beq
\alpha n_{\rm H}x^2 \simeq \epsilon _{\rm i}H.
\eeq
The definition of the parameter $\epsilon _{\rm i}$ follows
equation~(\ref{eq:source}), and $\alpha$ is the recombination
coefficient for principal quantum numbers $n\geq 2$. At this
ionization the optical depth for Thomson scattering is
\beq
\tau \sim \epsilon _{\rm i}^{1/2}
	\left( {1-Y_{p} \over 0.76} \right)^{1/2}
        \left( { \Omega_{\ba} h^{2} \over 0.02 } \right)^{1/2}
	\left( { \Omega_{\mat}h^{2} \over 0.122} \right)^{-1/4}
	\left( z \over 10^{3} \right)^{1.2}.
\label{eq:tau_i}
\eeq
Figure 2 shows in more detail the effect of the production of ionizing photons
on the ionization history.
If $\epsilon _{\rm i} \gsim 1$ the slow
decrease in the optical depth through $\tau\sim 1$ would remove
the first peak in the anisotropy power spectrum, an effect 
the measurements indicate is unacceptable. 

Sources of \lya photons may also add energy to the CMB by
inverse Compton scattering, but if the energy added is comparable
to the energy in \lya photons the effect on the $y$-parameter is
small.  

\section{Temperature Anisotropy Power Spectrum}
\label{sec:cl}
\begin{figure}
\plotone{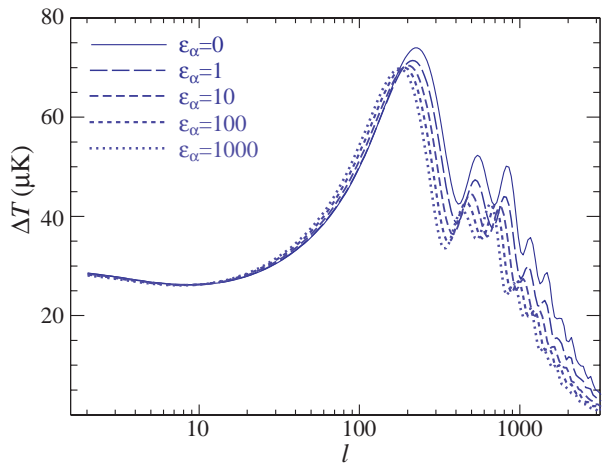}
\caption{The temperature anisotropy power spectrum ($\Delta T = 
[\ell(\ell+1)C_{\ell}/2\pi]^{1/2}$)
for various values of the \lya production parameter $\epsilon_{\alpha}$.}
\end{figure}

In the adiabatic CDM model the ionization history and the
cosmological parameters fix the CMB anisotropy power 
spectrum; the results in Figure~2 for the cosmologically flat
model with the parameters in Figure~1, and in Figure~3 for an
open model with the cosmological parameters $\Omega_\mat=0.6$,
$\Omega_{\Lambda}=0$, $h=0.7$, $\Omega_\ba h^{2}=0.02$
are computed using a code based on \cite{WhiSco96} (1996). 

The shift in the angular wavenumber
$l_1$ at the peak is noteworthy because $l_1$ is
used to infer the space curvature. A general expression for
its value in adiabatic models is
(\cite{HuSug95} 1995, with $\ell_1 \approx 0.73 \ell_p$)
\begin{eqnarray}
    \ell_{1} &\approx& {125 \over \sqrt{\Omega_\mat +\Omega_\Lambda}}
        [1+\ln(1-\Omega_\Lambda)^{0.085}]
        \left({z_* \over 10^3}\right)^{1/2} 
\nonumber
        ,\\
&& \times \left( {1 \over \sqrt{R_*}} \ln
        {\sqrt{1+R_*} +  \sqrt{R_* + r_* R_*}
        \over  1 + \sqrt{r_* R_*}} \right)^{-1}  \,,
\label{eqn:correction}
\end{eqnarray}
where 
\begin{eqnarray}
r_*      & = & 0.042 m^{-1} (z_*/10^3)\,, \nonumber\\
R_*      & = & 30 b (z_*/10^3)^{-1}\,, \nonumber\\
z_*(0) &\approx& 1008 (1+0.00124 b^{-0.74})(1+c_1 m^{c_2})\,,\nonumber\\
c_1    &=& 0.0783 b^{-0.24} (1+39.5 b^{0.76})^{-1}\,,\nonumber\\
c_2    &=& 0.56 (1+21.1 b^{1.8})^{-1} \,,
\end{eqnarray}
with $b\equiv\Omega_\ba h^2$ and $m \equiv \Omega_\mat h^2$. 	
Although in isocurvature models $\ell_{1}$ is larger by 50\% or more, 
it scales with cosmological parameters in the same way.

The leading order dependence of the peak scale is then
\begin{equation}
\ell_1 \propto (\Omega_\mat + \Omega_\Lambda)^{-1/2} z_{*}^{1/2}\,,
\end{equation}
such a $10\%$ change in $z_{*}$ shifts $\ell_1$ by 5\% and can
compensate a 10\% change in the spatial curvature
$\Omega_\mat + \Omega_\Lambda$.  In our model for
delayed recombination with $z_{*}$ given by eqn.~(\ref{eqn:shift}), 
$\epsilon_{\alpha}=1000$ can make an open universe with 
$\Omega_{\mat}=0.6$ appear flat (see Figure 4).

\begin{figure}
\plotone{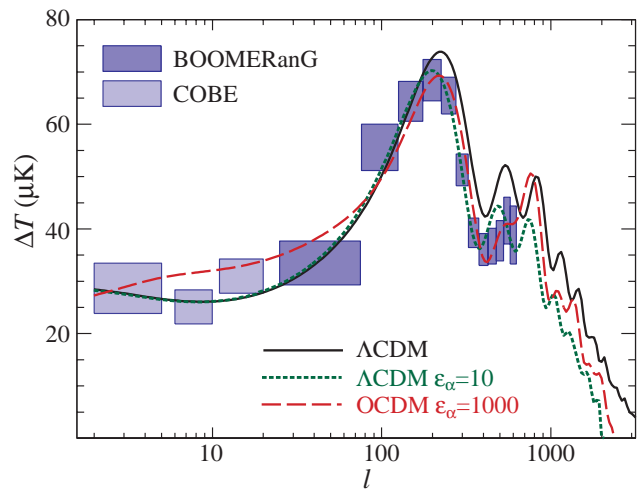}
\caption{A delay in recombination (with $\epsilon_{\alpha}=1000$) can 
make an open universe ($\Omega_m=0.6$, dashed line)  appear flat (solid line,
$\Omega_m=0.25$, $\Omega_\Lambda$=0.75) or decrease $\ell_1$ in a flat universe
and suppress the secondary peaks (with $\epsilon_{\alpha}=10$).  
The COBE and BOOMERanG data are plotted as bandwidth $\times 1\sigma$ error boxes.}
\end{figure}

Even in the context of cosmologically flat models, one might want
to consider the possibility of delayed recombination.  The observed
peak at $\ell_{1} = 197 \pm 12$ (95\% CL, \cite{deBer00} 2000) 
favors either a slightly closed low density universe or one of 
high density ($\Omega_{\mat} h^{2}$).  While flat models are certainly still
consistent with the data, the expected increase in precision of
the measurements could force one into accepting either a closed 
geometry, large Hubble constant or delayed recombination.


The amplitudes of the secondary peaks break
this approximate degeneracy of delayed recombination with space
curvature or the model for structure formation. While 
the promptness of the delayed recombination nearly preserves 
the first peak in the spectrum
the secondary peaks are strongly suppressed because of the extra
time the photons are allowed to diffuse (\cite{Sil68} 1968).  
As discussed in \cite{HuWhi96} (1996), the ratio of the first
peak location to the location of the damping tail is a sensitive
test of delayed recombination.   An equally important but more
subtle effect is that a delay in recombination raises the
baryon-photon momentum density ratio $R_*$ due to the redshifting
of the photons.  This effect suppresses the second peak and
raises the third.  
Though an increase in the baryon density parameter also has this
effect, it simultaneously reduces rather than increases the
damping and can in principle
be distinguished through the higher peaks.

Under standard recombination, the 
the ratio of power at the second
versus first peak scales as
\begin{equation}
A_2(0) \equiv {C_{\ell_{2}} \over C_{\ell_{1}}} 
	\approx 0.7\left[1+\left({\Omega_b h^2 \over 0.016}\right)^4\right]^{-1/4}
		2.4^{n-1}
\end{equation}
In our model, $A_2(\epsilon_\alpha) \approx A_2(0) z_*(\epsilon_\alpha)/z_*(0)$
such that delayed recombination is twice as effective in suppressing
power at the second peak as it is at shifting the first peak.
Observations currently favor $A_2 \sim 1/3$ (see Fig.~4)
compared with $\sim 0.5$ in our fiducial $\Lambda$CDM model with 
standard recombination.  Of course the tilt $n$ and the baryon density
can also lower this ratio.

Because of its cumulative effect over the secondary peaks,
$\epsilon_{\alpha}$ has statistically significant effects as long as 
it is greater than 
\begin{equation}
   \epsilon_{\rm min} = \left[\sum_{\ell=2}^{\ell_{\rm max}}(\ell+1/2)
    \left({\partial \ln C_{\ell} \over 
    \partial\epsilon}\right)^{2}\right]^{-1/2}\,,
\end{equation}
where $\ell_{\rm max}$ is the largest $\ell$ for which the measurements
are cosmic variance limited.  For $\ell_{\rm max}=500$, $\epsilon_{\rm 
min} = 0.006$; for $\ell_{\rm max}=1000$, $\epsilon_{\rm min}=0.002$.

\section{Discussion}
Our model for delayed recombination with $\epsilon _{\alpha} = 1$
yields a 5\% shift of the first peak and a reduction
of the secondary peak by 10\%, an observationally
interesting effect.
The sources must have $\epsilon _{\rm i} \ll 1$ 
(see Figure 2). This condition requires
that the \lya sources (perhaps hot stars or quasars) be
surrounded by envelopes of neutral primeval material
which strongly absorbs the ionizing radiation.

Our picture does not follow from the conventional CDM model for
structure formation. Apparent problems with excess  
small-scale clustering in this model
(\cite{Mooetal99} 1999; \cite{Klyetal99} 1999)
have motivated discussions of modifications 
(\cite{SpeSte99} 1999; \cite{KamLid99} 1999; \cite{Pee00} 2000;
\cite{HuBarGru00} 2000). 
These modifications
would tend to delay the appearance of the first generation
of gravitationally
bound systems, however, which is in the opposite direction to 
what is postulated here.
These lines of thought do not rule out early sources of \lya
photons, of course, but they do suggest one might best look for a
source outside the adiabatic CDM model. Perhaps cosmic strings
produced wakes that were subdominant to the primeval CDM density
fluctuations in determining the mass fluctuation power spectrum
but did produce occasional non-gaussian density fluctuations 
(\cite{ConHinMag99} 1999) large enough to have collapsed to
stars or active black holes that could have produced \lya
photons.  
 
To summarize, we have argued that the picture of delayed
recombination caused by early sources of \lya radiation, modeled
after quasar spectra, but with suppressed X-ray emission,
has the virtue that it naturally preserves
the observed first peak in the CMB temperature angular power
spectrum while shifting the peak to larger scales. The picture is
ad hoc but important as a conceivable complication in the
application of an exceedingly powerful cosmological test. The
most direct diagnostic for delayed recombination 
seems to be the suppression of the
secondary peaks in the spectrum. 
The amplitude of the third peak, which may be measured by the
BOOMERanG experiment or the MAP satellite\footnote{\tt http://map.gsfc.nasa.gov},
is crucial for distinguishing the effect of delayed recombination from
that of adjustments of the values of space curvature, the baryon density,
the Hubble constant or the shape of the spectrum of the primeval density
fluctuations.  If the measured 
secondary peaks agreed with the standard model for 
recombination with astronomically acceptable cosmological
parameters that fit the primary peak, it would convincingly rule
out our model for delayed recombination.

\acknowledgements
We have benefitted from discussions with Martin Rees.
P.J.E.P. is supported in part by the NSF. 
S.S. is supported by NSF grant PHY-9513835.
W.H. is supported by the Keck Foundation and a Sloan Fellowship.

\end{document}